\documentstyle[nato,epsf]{crckapb}
\begin{opening}
\title{Dislocation-free 3D islands in highly mismatched epitaxy: An
equilibrium study with anharmonic interactions}
\author{Ivan Markov}   
\institute{Institute of Physical Chemistry, Bulgarian Academy of Sciences,\\ 
1113 Sofia, Bulgaria}
\author{Jos\'e Emilio Prieto}
\institute{Departamento de F\'\i{}sica de la Materia Condensada and Instituto
Universitario de Ciencia de Materiales ``Nicol\'as Cabrera",\\Universidad
Aut\'onoma de Madrid, 28049 Madrid, Spain}

\end{opening}
\begin{document}
\begin{abstract}
Accounting for the anharmonicity of the real interatomic potentials in a
model in 1+1 dimensions shows that coherent 3D islands can be formed on the
wetting layer in a Stranski-Krastanov growth mode predominantly in compressed
overlayers. Coherent 3D islanding in expanded overlayers could be expected
as an exception rather than as a rule. The thermodynamic driving force of
formation of coherent 3D islands on the wetting layer of the same material
is the weaker adhesion of the atoms near the islands edges. The
average adhesion gets weaker with increasing island's thickness but
reaches a saturation after several monolayers. A misfit greater than a
critical value is a necessary condition for coherent 3D islanding.
Monolayer height islands with a critical size appear as necessary precursors
of the 3D islands. The 2D-3D transformation from monolayer-high islands to
three-dimensional pyramids takes place through a series of stable intermediate 
3D islands with discretely increasing thickness.

\end{abstract}

\section{Introduction}

Instabilities during growth of surfaces are of crucial importance for
fabrication of devices~\cite{Politi}. Of particular interest in recent time
is the
instability of the two-dimensional (2D) layer-by-layer growth against the
formation of coherently strained (dislocation-free) three-dimensional (3D)
islands of nanometer scale in highly mismatched epitaxy. The latter is known
as a ``coherent Stranski-Krastanov" (SK) growth~\cite{Eagle}, and is a subject
of intense research owing to possible optoelectronic
applications as lasers and light emiting diodes~[2-4]. That is why much effort
has been made in the last decade to determine the
equilibrium shape of the crystallites as a function of the
volume~[5-8], the change of shape during
growth~\cite{DarTer}, the strain distribution within the coherent islands and
their energy~[10-20], the kinetics of growth of arrays of quantum dots and the
physical reason of the narrow size distribution~[21-23], which is often
experimentally observed.

The physical reason of occurrence of the Stranski-Krastanov growth mode is
generally inderstood. Too much strain energy accumulates into the film during
the initial planar growth, and the strong adhesion exerted by the substrate
(which is the reason for the planar growth) disappears beyond
several atomic diameters. A wetting layer (WL) composed of an integer number
of equally strained monolayers is thus formed. The growth continues further
by the formation of 3D crystallites, in which the additional surface energy
is overcompensated by the strain relaxation. In other words, {\it the higher
energy phase representing a homogeneously strained planar film is replaced
beyond some critical thickness by a lower energy phase of (completely or
partially) relaxed 3D crystallites.}

Although the essential physics seems clear, too many questions of fundamental
character remain to be answered. As the atoms on top of the surface of the
wetting layer do not ``feel" energetically the presence of the substrate and
both the wetting layer and the 3D islands consist of one and the same
material, we can consider as a first approximation the formation of coherent
3D clusters in SK growth as {\it homoepitaxial growth} on an uniformly
strained crystal surface. If so, it is not clear what is the thermodynamic
driving force for 3D islanding if the islands are coherently strained to the
same degree as the underlying wetting layer. This question is closely
connected with the structure and energy of the boundary between the 3D
islands and the wetting layer. The energy of this boundary is often taken
equal to zero~\cite{TerTr}. This means a complete wetting of the 3D islands
by the substrate (the WL) which rules out the 3D islanding from a
thermodynamic point of view. It is also not clear why coherent 3D islands
are observed in compressed rather than in expanded overlayers, and at values
of the misfit $\varepsilon _{0} = \Delta a/a$ that are huge for materials
with directional and brittle covalent bonds (InAs/InP (3.2\%)~\cite{Rud},
Ge/Si (4.2\%)~\cite{Eagle,Mo}, InAs/GaAs (7.2\%)~\cite{Moison}, CdSe/ZnSe
(7.6\%)~\cite{Schik}). The only exception, to the authors' knowledge, of
expanded overlayer, is the system PbSe/PbTe (-5.5\%)~\cite{Pinc}. Other
question is whether the misfit should be greater than some critical value in
order for the
coherent 3D islanding to take place. Are two-dimensional monolayer height 
islands necessary precursors for the formation of 3D islands as suggested 
by some authors~\cite{Lannoo,Chen,StMar,Elka}? If yes, is there a critical 
volume size
(or a size of the 2D island) for the 2D-3D transformation to occur? What is
the pathway of the latter, does it pass through a series of intermediate
states with increasing thickness, and are these states stable or metastable?
In this paper we make an attempt to answer at least qualitatively some
of the questions posed above.

The thermodynamic driving
force for occurrence of one or another mode of growth should be given by the
difference $\Delta \mu  = \mu (n) - \mu ^0_{3D}$ of the chemical potentials
$\mu (n)$ of the film, and $\mu ^0_{3D}$ of the bulk crystal of the same
material. The film chemical potential depends on the thickness measured in
number $n$ of monolayers owing to the thickness distribution of the misfit
strain and the attenuation of the energetic influence of the
substrate~[31-33]. If we deposit a crystal $A$ on the
surface of a crystal $B$ the thermodynamic driving force can be written in
terms of interatomic energies $\Delta \mu  = E_{AA}{\it \Phi }$ where
${\it \Phi } = 1 - E_{AB}/E_{AA}$ is the so-called adhesion parameter which
accounts for the wetting of the substrate by the overgrowth~\cite{Kaisch}.
$E_{AA}$ and $E_{AB}$ are the energies per atom to disjoin a half-crystal
$A$ from a like half-crystal $A$ and from an unlike half-crystal $B$,
respectively. $E_{AB}$ is in fact the adhesion energy which includes in
itself the
thickness distribution of the strain energy due to the lattice misfit, and
the attenuation of the bonding with the substrate~\cite{Kern,Mark1}. The
adhesion parameter $\it \Phi $ is the same which accounts for the influence
of the substrate on the work of formation of 3D nuclei of different material
on top of it in the classical nucleation theory~\cite{Mark1}.
Replacing the bonding energies $E_{AA}$ and $E_{AB}$ by the corresponding
surface energies gives the famous 3-$\sigma $ criterion of Bauer for the mode
of growth $\Delta \mu  = a^2[\sigma _A + \sigma _{AB}(n) -
\sigma _B]$~\cite{Bauer}, where $a^2$ is the area occupied by an atom at the
interface.

\begin{figure}[htb]
\centering{\epsfysize=6cm \epsffile{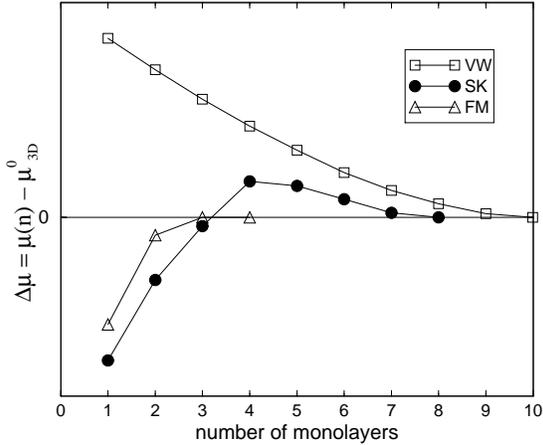}}
\caption{\label{modenew}
Schematic dependence of the thermodynamic driving force $\Delta \mu  =
\mu (n) - \mu ^0_{3D}$ which determines the occurrence of a given mode of
growth on the film thickness in number of monolayers: VW - Volmer-Weber,
SK - Stranski-Krastanov, and FM - Frank-van der Merwe. Note that in the
case of FM growth the points denote the chemical potentials of the separate
monolayers as only the uppermost incomplete monolayer determines the
equilibrium vapor pressure. In the other extreme of VW growth,
all monolayers are incomplete and the chemical potential will be given by
the mean value of the chemical potentials of all constituent monolayers.
This fact was realized by Stranski and Krastanov themselves in their seminal
paper (Sitzungsber. Akad. Wissenschaft Wien {\bf 146}, 797 (1938), see for
review Refs. [33,35]).}
\end{figure}
The thickness dependence of the film chemical potential is schematically
illustrated in Fig.\ \ref{modenew}. In the two limiting cases of
Volmer-Weber (VW) (incomplete wetting, $0 < {\it \Phi } < 1$) and Frank-van
der Merwe (FM) growth (complete wetting, ${\it \Phi } \le 0, \varepsilon _{0}
\cong 0$) $\mu (n)  - \mu ^0_{3D}$ goes asymptotically to zero from above
and from below, respectively, but changes sign in the case of SK growth
(${\it \Phi } \le 0, \varepsilon _{0} \neq 0$)~\cite{Kern,Stoyan,Mark1,Gilmer}.
In the latter case, beyond the maximum, we consider the 3D islands as the
overlayer material $A$, and the wetting layer as the substrate crystal $B$.
Thus the strained wetting layer and the relaxed 3D islands represent
necessarilly different phases in the sense of Gibbs. The wetting layer can be
in equilibrium only with an {\it undersaturated vapor phase}, whereas the 3D
islands are in equilibrium with a {\it supersaturated vapor}. The dividing
line is $\Delta \mu  = 0$ at which the wetting layer cannot grow thicker and
the 3D islands cannot nucleate and grow. Thus the adhesion parameter
${\it \Phi } = \Delta \mu  /E_{AA}$ relative to the cohesion energy $E_{AA}$
is in fact equal to the thermodynamic driving force for the occurence of one
or another mode of growth. In other words, we can treat the SK mode as a FM
mode driven by complete wetting ($\Delta \mu  < 0$), followed by VW mode
driven by incomplete wetting ($\Delta \mu  > 0$). (The more rigorous
definition is $d\mu /dn < 0$ or $d\mu /dn > 0$~\cite{Stoyan}). The question is
how the lattice misfit can lead to incomplete wetting (${\it \Phi } > 0$) on
the surface of the wetting layer if the energetic influence exerted by the
substrate is already lost, i.e. $E_{AB} \rightarrow E_{AA}$. In the classical
SK mode the incomplete wetting is due to the introduction of misfit
dislocations (MDs)~\cite{Matt}. Once we know the answer of this question in
the case of a coherent SK growth we could easily find the answers of the
others.

\section{Model}

We consider an atomistic model in $1+1$ dimensions (substrate + height) which
we treat
as a cross section of the real $2+1$ case. An implicit assumption is that in 
the real $2+1$ model the monolayer islands have a compact rather than a
fractal shape and the lattice misfit is one and the same in both orthogonal
directions. Furthermore, we exclude from our considerations the possible
interdiffusion and the subsequent gradient of strain as found
recently by Kegel {\it et al}~\cite{Kegel} in the case of InAs/GaAs quantum
dots. The 3D islands are represented by linear chains of atoms stacked
one upon the other as in the model proposed by Stoop and van der
Merwe~\cite{Stoop} and later by Ratsch and Zangwill~\cite{Ratsch}, each upper
chain being shorter than the lower one by one atom. In this sense the lateral
size, and particularly the height of the islands are {\it discrete}
parameters, whereas in most of the theoretical considerations they are taken
as continuous variables~\cite{Politi,Duport,TerTr}.

In a previous paper~\cite{Elka}, we used the method of computation proposed by
Ratsch and Zangwill~\cite{Ratsch}, which is based on the well-known model of
Frenkel and Kontorova~\cite{Frenkel}. The latter treats the overlayer as a 
linear chain of atoms subject to an external periodic potential exerted by a
rigid substrate~\cite{Frenkel,FM}. Ratsch and Zangwill accepted that each
layer (chain) presents a rigid sinusoidal potential to the chain of atoms on
top of it. The potential trough separation of the lower chain is taken
constant and equal to the average of all trough separations. As the strains
of the bonds that are closer to the free ends are smaller, the average bond
strain of each upper chain is closer to zero. In other words, the lattice
misfit decreases from $\varepsilon _{0}$ at the island's base to zero at the
apex. This method is, however, inadequate to
describe properly a thickening overlayer because of one basic assumption,
namely, the rigidity of each monolayer upon formation of the next one on top
of it. This assumption rules out the relaxation and redistribution of the
strains in the lower layers when upper layers are added. In particular,
this method does not allow to compute the structure and energy of the
interfacial boundary between the wetting layer and the 3D islands upon
thickening of the latter.

For the above-mentioned reasons, in the present work we make use of a
simple minimization procedure.
The atoms interact through a Morse potential that can be easily generalized
to vary its anharmonicity by adjusting two constants $\mu $ and $\nu $
($\mu  > \nu $) that govern separately the repulsive and the attractive
branches, respectively~[44-46],
\begin{eqnarray}\label{potent}
V(x) = V_{o}\Biggl[\frac{\nu }{\mu - \nu }e^{-\mu (x-b)} - \frac{\mu }{\mu - 
\nu }e^{-\nu (x-b)}\Biggr],
\end{eqnarray}
where $b$ is the equilibrium atom separation. For $\mu  = 2\nu $ the
potential (\ref{potent}) turns into the familiar Morse potential,
which has been used in the present work for the case $\nu = 6 $.

The pair potential designed by Tersoff for description of the properties of
materials with directional covalent bonds like Si contains an additional
parameter which accounts for the local atomic environments around the
neighboring atoms~\cite{Jerry}. He showed that most of the properties of Si
could be computed with an error smaller than 1\%, compared with experimental
data and {\it ab initio} calculations, by accounting only for the first
neighbor interactions. For this reason,  we occasionally consider only 
interactions in the first coordination sphere in order to mimic the
directional bonds that are characteristic for the most semiconductor
materials.

Our programs calculate the interaction energy of all the atoms as well as its
gradient with respect to the atomic coordinates, i.e. the forces. Relaxation
of the system is performed by allowing the atoms to displace in the direction
of the gradient in an iterative procedure until the forces fall below some
negligible cutoff value. The calculations were performed under the assumption
that the substrate (the wetting layer) is rigid. This assumption is strictly
valid in the beginning of the 2D-3D transformation when the 3D islands are
still very thin~\cite{Voltaire}.

Yu and Madhukar computed recently, by making use of the Stillinger-Weber
interatomic potential~\cite{SW} in a molecular dynamics study, the
distribution of the strains and stresses in and around a 3D Ge island having
a shape of a full pyramid with a length of the base edge 326 {\AA} and a
heigth of 23 monolayers~\cite{Yu}. They found that the atoms in the middle of
the first atomic plane of a coherent 3D Ge island are displaced upwards by
0.6{\AA} that is approximately half of the interplanar spacing of Ge(001)
(1.4\AA), whereas the atoms at the island's edges are displaced slightly
downwards. The same holds for the vertical displacements of the atoms
belonging to the uppermost Si plane. As the vertical displacements strongly
influence the adhesion of the islands to the wetting layer we also performed
preliminary calculations in which the uppermost three monolayers were allowed
to relax. The results of these calculations demonstrated qualitatively the
same behavior as in the case of a rigid substrate. For this reason, we present
here only the results obtained under the assumption of the rigid substrate.
Detailed systematic studies of the effect of the substrate relaxation will be
published elsewhere.

\begin{figure}
\epsfxsize=12.5cm \epsffile{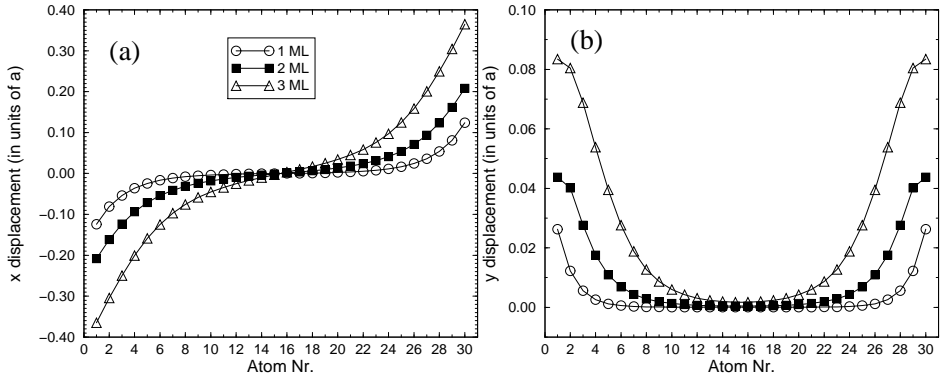}
\caption{\label{displ}
Horizontal (a) and vertical (b) displacements of the atoms of the base chain 
from the bottoms of the potential troughs provided by the homogeneously
strained wetting layer for a misfit of 7\%. The displacements are given 
in units of a, the lattice parameter of the substrate and wetting layer. 
They increase with increasing island's thickness taken in number of 
monolayers. Islands of 30 atoms in the base chain were considered.}
\end{figure}

\section{Results}

Fig.\ \ref{displ}(a) shows the horizontal displacements of the atoms of the 
base chain of a coherently strained island, for a value of the misfit of 7\%. 
The displacements are referenced to the sites the atoms would occupy if they 
belonged to the next complete monolayer, which would then be a part of 
the wetting layer. It can be 
seen that the end atoms are strongly displaced as in the model of Frenkel and
Kontorova~\cite{Frenkel} and of Frank and van der Merwe~\cite{FM}. Increasing
the island height leads to greater displacements of the end atoms. The reason
is the effective increase of the strength of the lateral interatomic bonding
in the overlayer with greater thickness as predicted by
van der Merwe {\it et al.}~\cite{Voltaire}. According to these authors an
island with a bilayer height could be approximately simulated by a monolayer
height island but with twice stronger lateral bonds. Fig.\ \ref{displ}(b)
shows the vertical displacements of the base atoms relative to the
interplanar spacing between the monolayers belonging to the wetting layer.
It is obvious that the vertical displacements are due to the climbing of the
atoms on top of underlying atoms as a result of the horizontal displacements.
The thicker the islands the greater are the horizontal displacements (for
reasons discussed above) and in turn the vertical displacements.
The results shown in Fig.\ \ref{displ} clearly demonstrate that the bonds
that are close to the island's edges are much less strained compared with
these in the middle, in agreement with the results obtained by Ashu and
Matthai~\cite{Ashu} and Orr {\it et al.}~\cite{Orr} but contrary to the
finding of Yu and Madhukar~\cite{Yu}.

\begin{figure}
\centering{\epsfysize=6cm \epsffile{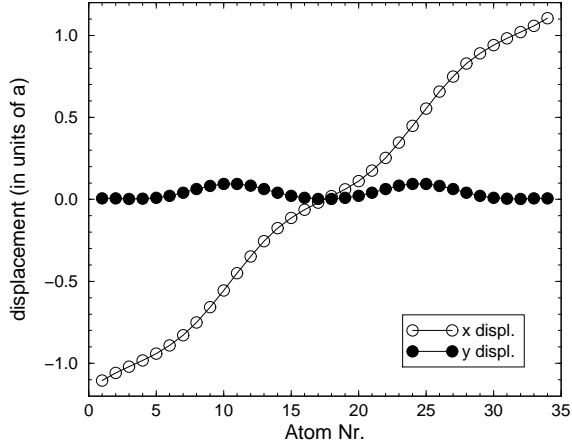}}
\caption{\label{mds}
Horizontal (x) and vertical (y) displacements of the atoms of the base chain
of an island three monolayers thick and containing two MDs.
The island contains a total amount of 99 atoms (34 in the base chain) and
the lattice misfit is 7\%.}
\end{figure}

\begin{figure}
\centering{\epsfysize=6cm \epsffile{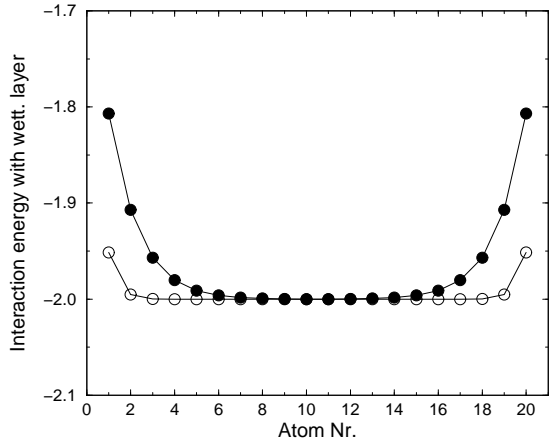}}
\caption{\label{phin}
Distribution of the energy (in units of $V_{o}$) of first-neighbors 
interaction, $E_{AB}(n)$, between the atoms of the base chain ($A$) 
of a monolayer-high, coherent island consisting of 20 atoms, and the 
underlying wetting layer, $B$, for positive ($\bullet$) and negative 
($\circ$) misfits of absolute value 7\%.}
\end{figure}

\begin{figure}
\centering{\epsfysize=6cm \epsffile{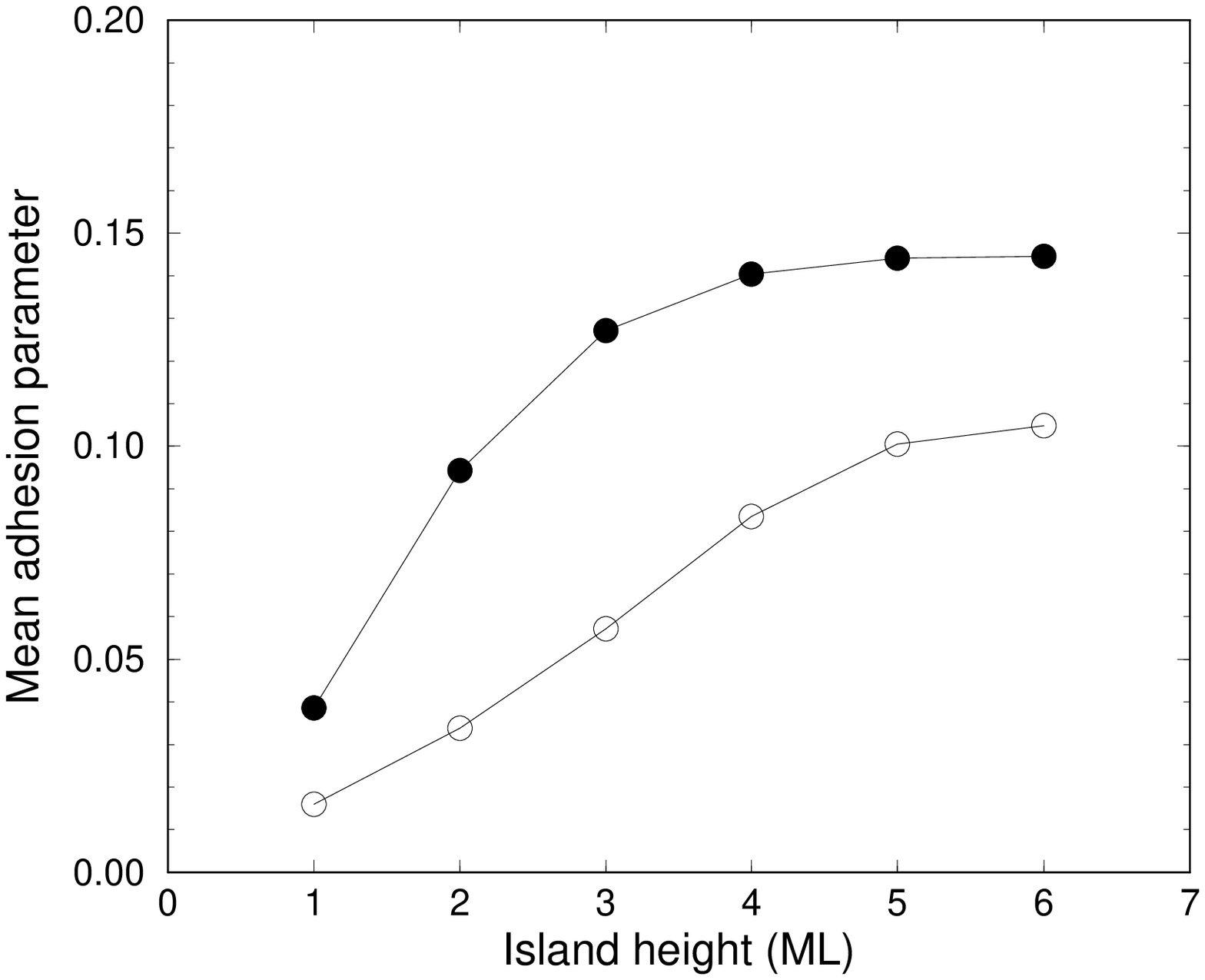}}
\caption{\label{phi}
Mean adhesion parameter $\it \Phi $ as a function of the islands' height in
number of monolayers for positive ($\bullet$) and negative ($\circ$) values
of the misfit of absolute value of 7\%. Coherent islands of 14 atoms in 
the base chain were considered in the calculations.}
\end{figure}

The interconnection between the vertical and the horizontal displacements is
beautifully demonstrated in Fig.\ \ref{mds} where they are shown in an
island containing two MDs. The horizontal displacements in
this case are greatest in the cores of the MDs and so are the vertical
displacements. This figure shows in fact the physical reason for the
incomplete wetting in the classical SK mode. The adhesion is weaker owing to
the introduction of MDs.

In order to illustrate the effect of the atom displacements on the adhesion
of the separate atoms belonging to the island's base chain,  we plot their
energy of interaction with the underlying wetting layer (Fig.\ \ref{phin})
for coherently strained islands.
As seen the atoms that are near to the chain ends (island's edges) adhere
much more weakly with the substrate. The influence of the potential
anharmonicity is clearly demonstrated. Only one or two end atoms in the
expanded chain adhere more weakly to the substrate whereas more than half of
the atoms at both ends in the compressed chain are weakly bound. The figure
demonstrates in fact the physical reason for the coherent SK mode which is
often overlooked in theoretical models. Moreover, it is a clear evidence 
of why compressed rather than expanded overlayers exhibit greater tendency 
to coherent SK growth.

It follows from Fig.\ \ref{displ} that increasing the island thickness leads to
weaker adhesion of the 3D islands to the wetting layer and, in turn, to the
stabilization of the coherent 3D islands. This is clearly demonstrated in
Fig.\ \ref{phi} which shows the
dependence of the mean adhesion parameter $\it \Phi $ on the islands' height
for positive and negative values of the misfit. 
It is calculated as the average of the interaction energy between the
base chain atoms and those of the wetting layer and is referenced to 
the corresponding value for a non-misfitting monolayer of the same size. 
It can be seen that $\it \Phi $ saturates beyond a thickness of about 
5 monolayers as expected. Note that in this case the
incomplete wetting ($\it \Phi >$ 0) is due solely to the misfit, the bonding
in both phases $A$ (the wetting layer) and $B$ (the 3D islands) being nearly
one and the same. What is more
important is that the adhesion parameter in compressed islands is visibly
larger than that in expanded islands which is due to the anharmonicity of the
interatomic potential. This behavior clearly shows the greater tendency of
the compressed overlayers to form coherent 3D islands. Another important
feature that
characterizes the mean adhesion parameter is its large absolute value. It is
comparable with the values that lead to 3D islanding in VW mode of growth on
chemically unlike surfaces~\cite{Mark1}.

\begin{figure}
\centering{\epsfysize=7cm \epsffile{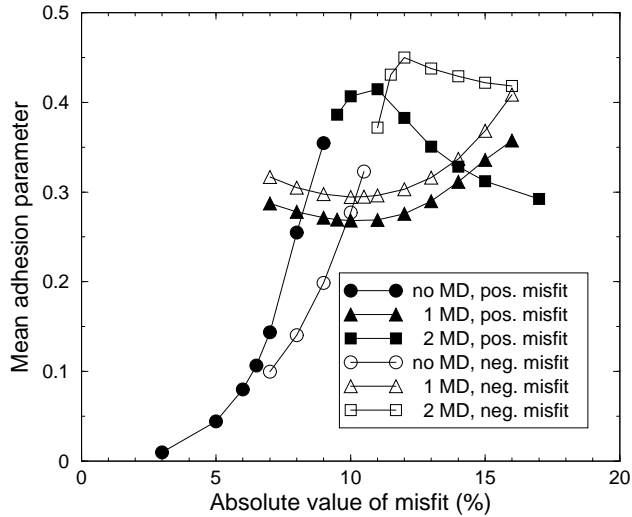}}
\caption{\label{phif}
Mean adhesion parameter as a function of the lattice misfit. The points 
correspond to the saturated values from curves as those shown in Fig. 5 
for coherent islands and, in addition, for  islands containing one and two 
MDs. The islands contain 14 atoms in the base chain
and have a height of 5 ML. Data for both positive and negative misfits are 
shown in one quadrant for easier comparison.}
\end{figure}

Fig.\ \ref{phif} shows the mean adhesion parameter $\it \Phi $ as a function
of the misfit both negative and positive for coherent islands as well as for
islands containing one and two MDs. As discussed in the Introduction this is
in fact the thermodynamic driving force for 3D islanding. Several interesting
properties are observed. First, the mean adhesion parameter of compressed
coherent islands is greater than that of expanded islands. This means that
the thermodynamic driving force for coherent 3D islanding is greater in
compressed rather than in expanded overlayers. In the absence of MDs the
incomplete wetting is due to the displacements of the end atoms (see Fig.\ 
\ref{displ}). In expanded overlayers the end atoms interact with their
neighbors by the weaker attractive branches and {\it vice versa}. As a result
$\it \Phi ^+_0 > \it \Phi ^-_0$.  

On the other hand, the opposite is observed for dislocated islands. This is
very easy to understand bearing in mind that in the classical (dislocated)
SK mode $E_{AB} \approx E_{AA} - E_{MD}$, and $\it \Phi  \approx
E_{MD}/E_{AA}$~\cite{Mark1}, $E_{MD}$ being the energy per atom of the MDs.
MDs have higher energy in expanded overlayers as they
represent regions with higher density of atoms which repulse each other with
the stronger repulsive branches of the potential. It is exactly the opposite
in compressed overlayers, so that $E_{MD}^{+} < E_{MD}^{-}$ and $\it
\Phi ^+_{MD} < \it \Phi ^-_{MD}$. This means that in the classical SK growth
the thermodynamic driving force of formation of dislocated 3D islands is
greater in expanded rather than in compressed overlayers.

Another property is that the
adhesion parameter of islands containing two MDs appears as a continuation
of that of the dislocation-free islands. This is also easily understandable
having in mind the similarity of the model with that of Frank and van der
Merwe~\cite{FM}. Dislocation-free solutions exist until the misfit reaches the
so-called {\it metastability limit} at which the end atoms reach the crests
between the next potential troughs and two dislocations (because of the
symmetry of the model) are simultaneously introduced at both free ends. The
energetic barrier for this process is equal to zero (for a review see Ref.
(\cite{Mark1}).

\begin{figure}
\centering{\epsfxsize=12.5cm \epsffile{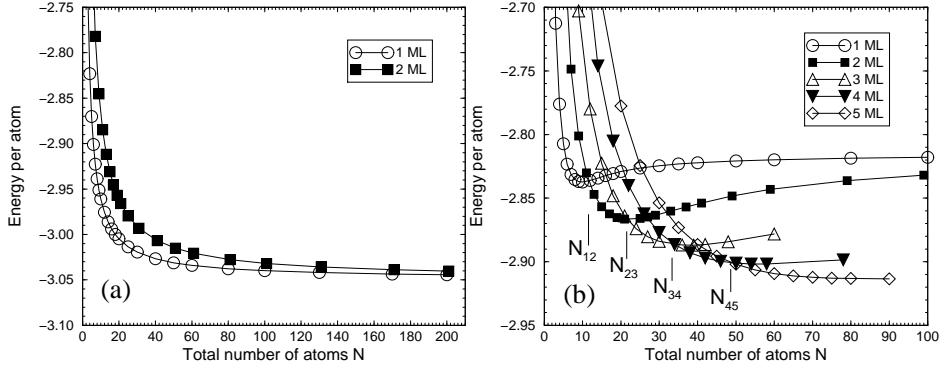}}
\caption{\label{e1234}
Dependence of the total energy per atom, in units of $V_{o}$, on the total 
number of atoms in compressed, coherently strained islands of different 
thicknesses, for two different values of the misfit: 
(a) $\varepsilon _{0} = 0.03$, (b)
$\varepsilon _{0} = 0.07$. The numbers $N_{12}$, $N_{23}$, etc. give the
limits of stability of monolayer, bilayer, ... islands, respectively. }
\end{figure}

In order to answer the questions posed in the Introduction we compare the
energies per atom of mono- and multilayer islands (frustums of pyramids) with
different thickness varied by one monolayer. The pyramids are bounded with
the steepest (60$^{\circ}$) sidewalls as they have the lowest energy
in models in $1+1$ dimensions~\cite{Ratsch,Elka}. As calculated, the energy 
represents a sum of the strain energy and the energy of the surfaces relative
to the energy of the same number of atoms in the bulk
crystal~\cite{StMar,Elka}.  Fig.\ \ref{e1234}(a) demonstrates
the energies per atom vs the total number of atoms of monolayer and bilayer
height islands at $\varepsilon _{0} = 0.03$. As seen the monolayer height
islands are always stable against the bilayer islands. The latter means that
the thermodynamics do not favor coherent 3D islanding. Monolayer height
islands will grow and coalesce until they cover the whole surface. MDs will
be then introduced to relieve the strain. Fig.\ \ref{e1234}(b)
demonstrates the same dependence (including also thicker islands) but at
larger value of the misfit $\varepsilon _{0} = 0.07$. This time the behavior
is completely different. The monolayer islands are stable against the bilayer
islands only upto a critical volume $N_{12}$, the bilayer islands are stable
in turn against the trilayer islands upto a second critical volume $N_{23}$,
etc. This behavior is precisely the same as in the case of VW growth where
the interatomic forces (the wetting) predominate and the lattice misfit plays
an additional role~\cite{StMar}. The same result (not shown) has been obtained
in the case of expanded overlayers ($\varepsilon _{0} < 0$) with the only
exception that monolayer height islands are stable against multilayer islands
upto much larger absolute values of the misfit.

\begin{figure}
\centering{\epsfysize=6cm \epsffile{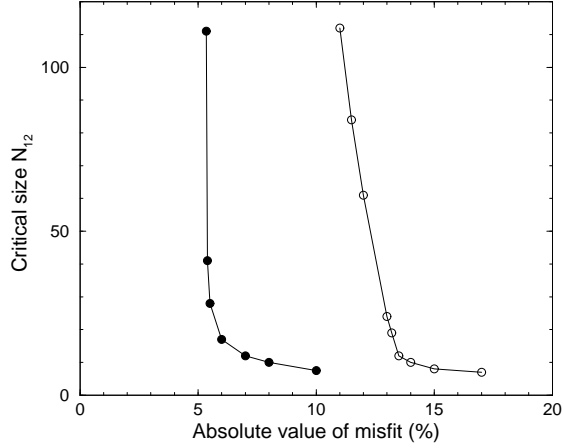}}
\caption{\label{N12}
Misfit dependence of the critical size $N_{12}$ (in number of atoms) for 
positive ($\bullet$) and negative ($\circ$) values of the lattice misfit. 
The curves are shown in one quadrant for easier comparison.}
\end{figure}

The mono-bilayer transformation is the first
step of the complete 2D-3D transformation. Studying the critical size
$N_{12}$ as a function of the misfit (see Fig.\ \ref{N12}) shows the
existence of critical misfits beyond which the formation of multilayer
islands can only take place. Below the critical misfit 
the monolayer height islands are stable irrespective of their size and the
growth will continue in a layer-by-layer mode until MDs are
introduced to relax the strain. The nearly twice larger
absolute value of the negative critical misfit is obviously due to the
anharmonicity of the atomic interactions. The weaker attractive interatomic
forces lead to smaller displacements both lateral and vertical of the end
atoms and in turn to stronger adhesion. The latter requires larger misfit in
order for the 3D islanding to take place.

\section{Discussion}

The existence of critical misfit clearly shows that the origin of the 3D
islanding in the coherent SK growth is the incomplete wetting which is due
to the atomic displacements near the islands edges. As seen in Fig.\ 
\ref{phif} the mean adhesion parameter $\it \Phi $, or which is the same,
the thermodynamic driving force $\Delta \mu $ for coherent
3D islanding has practically the same values as that in the case of the
classical (dislocated) SK mode at sufficiently large values of the misfit.
Moreover, the comparison of Fig.\ \ref{displ} and Fig.\ \ref{mds}  
shows that at a given misfit and a given thickness, there is a critical 
lateral size (or a critical volume) beyond which
MDs are spontaneously introduced to relieve the strain. 
It in fact determines the transition from the coherent to the classical
(dislocated) SK growth which in the real case should be accompanied with the
change of the shape. All the above leads to the conclusion that the physical
reason for both the classical and coherent SK mode is one and the same.

The average adhesion depends strongly on the anharmonicity of 
the interatomic forces. Expanded islands adhere more strongly to the wetting 
layer and the critical misfit beyond which coherent 3D islanding is possible 
is much greater in absolute value compared with that in compressed overlayers. 
As a result coherent SK growth in expanded films could be expected at very 
(unrealistically) large absolute values of the negative misfit. The latter, 
however, depends on the materials parameters (degree of anharmonicity, 
strength of the chemical bonds, etc.) of the particular system and cannot be 
completely ruled out. Xie {\it et al}~\cite{Xie} studied the deposition of 
Si$_{0.5}$Ge$_{0.5}$ films in the whole range of 2\% tensile to 2\% 
compressive misfit on relaxed buffer layers of Si$_{x}$Ge$_{1-x}$ starting 
from $x = 0$ (pure Ge) to $x = 1$ (pure Si). They found that 3D islands are 
formed only under compressive misfit larger than 1.4\%. Films under tensile 
misfit were thus stable against 3D islanding in excellent agreement with the 
predictions of our model.

The existence of a critical misfit for 2D-3D transformation to occur both in 
compressed and expanded overlayers has been noticed in practically all
systems studied so far. Pinczolits {\it et al.}~\cite{Pinc} have found that
deposition of PbSe$_{1-x}$Te$_{x}$ on PbTe(111) remains purely two
dimensional when the misfit is less than 1.6$\%$ in absolute value (Se
content $< 30\%$). Leonard {\it et al.}\cite{Leo} have successfully grown
quantum dots of In$_{x}$Ga$_{1-x}$As on GaAs(001) with $x = 0.5$
($\varepsilon _{0} \approx 3.6\%$) but 60\AA{} thick 2D quantum wells at
$x = 0.17$ ($\varepsilon _{0} \approx 1.2\%$). Walther {\it et
al.}~\cite{Walther} found that the critical In content is
approximately $x = 0.25$, or $\varepsilon _{0} \approx 1.8\%$. 
As commented before, a critical misfit of 1.4\% has been found by 
Xie {\it et al} upon deposition of 
Si$_{0.5}$Ge$_{0.5}$ films on relaxed buffer layers of Si$_{x}$Ge$_{1-x}$ 
with varying composition~\cite{Xie}.

A rearrangement of monolayer height (2D) islands into multilayer (3D) islands 
has been reported by Moison {\it et al.}~\cite{Moison} who established that the 
3D islands of InAs begin to form on GaAs at a coverage of about 1.75 ML but
then the latter suddenly decreases to 1.2 ML. This decrease of the coverage
in the second monolayer could be interpreted as a rearrangement of an amount
of nearly half a monolayer into 3D islands. The same phenomenon has been
noticed by Shklyaev, Shibata and Ichikawa in the case of
Ge/Si(111)~\cite{Ichi}.
Voigtl\"ander and Zinner noted that Ge 3D islands in Ge/Si(111) epitaxy have 
been observed at the same locations where 2D islands locally exceeded the 
critical wetting layer thickness of 2 bilayers~\cite{Voigt1}. Bhatti {\it et
al.}~\cite{Bhatti} and Polimeni {\it et al.}~\cite{Polimeni} also reported the
coexistence of large pyramids and small flat islands. These observations show
that the 2D islands really appear as precursors for the 3D islands.

The question of the existence and particularly the stability of the
intermediate states is more difficult to answer. Rudra {\it et al.} measured
photoluminescence (PL) spectra of InAs layers deposited on InP(001) at two
different temperatures (490 and 525$^\circ$C) and buried in the same
material~\cite{Rud}. When the layers were grown at 490$^\circ$C and the capping
layer was deposited immediately after the deposition of the InAs the spectrum
consisted of a single line. If the InAs layer was annealed for 10 s before
capping with InP the spectrum consisted of 8 lines. At 525$^\circ$C 3 lines
were observed already in absence of annealing. The above observations could
be explained by formation and coexistence of islands with different thickness
varying by one monolayer. Colocci {\it et al.}~\cite{Col} performed PL studies
of InAs deposits on GaAs(001) with thickness slightly varying around the
critical thickness of 1.6 monolayers for the onset of the 3D islanding. They
observed an increasing number of luminescence lines with increasing film
thickness. These lines were attributed to families of 3D islands with similar
shape but with heights differing by one monolayer. Flat platelets, 2 - 6
monolayers high, have been observed during the growth of GaN/AlN
heterostructures~\cite{Bourret}.

Although the above results seem to be in an excellent qualitative agreement
with the theoretical predictions of the model, the thermodynamic stability of
islands with quantized height of one monolayer, and the existence of a
critical misfit is still debated~\cite{Politi,Duport}. The reason of the
discrepancy of our results with those of Duport {\it et al.}~\cite{Duport}
most probably stems from the implicit assumption, made by the above
authors, that the widths of the lower, $R$, and the upper, $R^{\prime}$,
bases, and particularly
the height $h$, of the crystal having a shape of a frustum of a pyramid, they
consider, are continuous variables. This is correct if the crystals are
sufficiently large. However, the continuum approximation is not acceptable in
the beginning of the 2D-3D transformation when the islands are still very
small (and thin). It is also not applicable in the limit $h \ll R$ for the
same reasons. The question of existence of a critical misfit follows
logically if we accept
that the intermediate states with heights differing by one monolayers exist
and are thermodynamically stable in consecutive intervals of the volume. 

In conclusion, accounting for the anharmonicity of the real interatomic
potentials in a model in $1+1$ dimensions, we have shown that coherent 3D
islands can be formed on the wetting layer in the SK mode predominantly in
compressed overlayers. Coherent 3D islanding in expanded overlayers could be
expected as an exception rather than as a rule. The thermodynamic driving
force for 3D islanding on the wetting layer of the same material is 
identified as the weaker adhesion of the atoms near the islands edges. 
This should also facilitate the 2D-3D transformation. 
Overcoming a critical lattice misfit appears as a necessary 
condition for the formation of coherent 3D islands. 
Monolayer height islands of a critical size appear as necessary
precursors of the 3D islands. The 2D-3D transformation from monolayer islands
to 3D pyramids takes place through a series of intermediate states with
heights increasing by one monolayer. The intermediate states are
thermodynamically stable in consecutive intervals of the volume. 

\acknowledgements

The authors are indebted to the Instituto
Universitario de Ciencia de Materiales ``Nicol\'as Cabrera" 
for granting research visits which enabled scientific collaboration.
This work was supported by the Spanish CICyT through project
Nr. MAT98-0965-C04-02.

%
%
\end{document}